%% file: main.tex
\documentclass[runningheads]{llncs}
\usepackage[disable]{todonotes}
\usepackage{graphicx}
\usepackage{arydshln}
\usepackage[T1]{fontenc}
\usepackage{listings}
\usepackage{amsmath,amssymb}
\usepackage[inline]{enumitem}
\usepackage{tikz}
\def\checkmark{\tikz\fill[scale=0.4](0,.35) -- (.25,0) -- (1,.7) -- (.25,.15) -- cycle;} 
\usepackage{caption}
\usepackage{subcaption}
\PassOptionsToPackage{hyphens}{url}

\lstdefinestyle{plogic}{
    basicstyle=\scriptsize\ttfamily,
    numbers=left,
    numberstyle=\scriptsize,
    stepnumber=1,
    numbersep=8pt,
    showstringspaces=false,
    breaklines=true,
    frame=lines,
    backgroundcolor=\color{background},
    language=sparql,
    xleftmargin=2em,
    framexleftmargin=2em,
    mathescape=true
}
\lstdefinestyle{sparql}{
    basicstyle=\scriptsize\ttfamily,
    numbers=left,
    numberstyle=\scriptsize,
    stepnumber=1,
    numbersep=8pt,
    showstringspaces=false,
    breaklines=true,
    frame=lines,
    backgroundcolor=\color{background},
    language=sparql,
    xleftmargin=2em,
    framexleftmargin=2em
}

\colorlet{punct}{red!60!black}
\definecolor{background}{HTML}{EEEEEE}
\definecolor{delim}{RGB}{20,105,176}
\colorlet{numb}{magenta!60!black}

\lstdefinelanguage{json}{
    basicstyle=\scriptsize\ttfamily,
    numbers=left,
    numberstyle=\scriptsize,
    stepnumber=1,
    numbersep=8pt,
    showstringspaces=false,
    breaklines=true,
    frame=lines,
    backgroundcolor=\color{background},
    literate=
     *{0}{{{\color{numb}0}}}{1}
      {1}{{{\color{numb}1}}}{1}
      {2}{{{\color{numb}2}}}{1}
      {3}{{{\color{numb}3}}}{1}
      {4}{{{\color{numb}4}}}{1}
      {5}{{{\color{numb}5}}}{1}
      {6}{{{\color{numb}6}}}{1}
      {7}{{{\color{numb}7}}}{1}
      {8}{{{\color{numb}8}}}{1}
      {9}{{{\color{numb}9}}}{1}
      {:}{{{\color{punct}{:}}}}{1}
      {,}{{{\color{punct}{,}}}}{1}
      {\{}{{{\color{delim}{\{}}}}{1}
      {\}}{{{\color{delim}{\}}}}}{1}
      {[}{{{\color{delim}{[}}}}{1}
      {]}{{{\color{delim}{]}}}}{1},
}
\renewcommand\tt[1]{\texttt{#1}}
\begin{document}
\title{Facade-X:\\an opinionated approach to SPARQL anything\\{\small Submitted to~\url{https://2021-eu.semantics.cc} (Accepted, May 2021)}}
\titlerunning{Facade-X: an opinionated approach to SPARQL anything}
%
\author{Enrico Daga\inst{1}\orcidID{0000-0002-3184-5407} \and
Luigi Asprino\inst{2}\orcidID{0000-0003-1907-0677} \and
Paul Mulholland\inst{1}\orcidID{0000-0001-6598-0757} \and
Aldo Gangemi\inst{3}\orcidID{}}
\authorrunning{E. Daga et al.}
%
\institute{The Open University (United Kingdom)
\email{\{enrico.daga,paul.mulholland\}@open.ac.uk}\and
University of Bologna (Italy) \email{luigi.asprino@unibo.it}\and
Consiglio Nazionale delle Ricerche (CNR) \email{aldo.gangemi@cnr.it}
}
\maketitle              
\begin{abstract}
\input{sections/abstract}
\end{abstract}
\section{Introduction}\label{sec:introduction}\vspace{-0.2cm}
\input{sections/introduction}

%
%
\vspace{-0.2cm}\section{Requirements}\label{sec:requirements}\vspace{-0.2cm}
\input{sections/requirements}

%
\vspace{-0.2cm}\section{An opinionated approach}\label{sec:approach}\vspace{-0.2cm}
\input{sections/approach}
\section{Implementation to SPARQL Anything}\label{sec:implementation}
\input{sections/implementation}
%
\vspace{-0.5cm}\section{Related work}\label{sec:relatedwork}
\input{sections/relatedwork}

\section{Evaluation}\label{sec:evaluation}
\input{sections/evaluation}
%
\vspace{-0.3cm}\section{Conclusions}\label{sec:conclusions}\vspace{-0.2cm}
\input{sections/conclusions}\vspace{-0.2cm}
%
%
%
\bibliographystyle{splncs04}
\bibliography{main}
\end{document}

%% file: sections/abstract.tex
The Semantic Web research community understood since its beginning how crucial it is to equip practitioners with methods to transform non-RDF resources into RDF. Proposals focus on either engineering content transformations or accessing non-RDF resources with SPARQL. Existing solutions require users to learn specific mapping languages (e.g. RML), to know how to query and manipulate a variety of source formats (e.g. XPATH, JSON-Path), or to combine multiple languages (e.g. SPARQL Generate). In this paper, we explore an alternative solution and contribute a general-purpose meta-model for converting non-RDF resources into RDF: \textit{Facade-X}. Our approach can be implemented by overriding the SERVICE operator and does not require to extend the SPARQL syntax. We compare our approach with the state of art methods RML and SPARQL Generate and show how our solution has
lower learning demands and cognitive complexity, and it is cheaper to implement and maintain,
while having comparable extensibility and efficiency.

%% file: sections/introduction.tex
Knowledge graphs have nowadays a key role in domains such as enterprise data integration and cultural heritage.
However, domain applications typically deal with heterogeneous data objects.
Therefore, ontology engineers develop knowledge graph construction pipelines that include the transformation of different types of content into RDF.
Typically, this is achieved by using tools that act as mediators between the data sources and the needed format and data model~\cite{haslhofer2011data}.
Alternatively, dedicated software components implement ad-hoc transformations from custom formats to a multiplicity of ontologies relevant to the domain~\cite{daga2016open}.
We place our research under the context of the EU H2020 SPICE project, which aims at developing a linked data infrastructure for integrating and leveraging museum collections using multiple ontologies covering sophisticated aspects of citizen engagement initiatives~\footnote{SPICE Project: \url{https://spice-h2020.eu}}.
Museum collections come in a variety of data objects, spanning from public websites to open data sets. 
These include metadata summaries as CSVs, record details as JSON files, and binary objects (e.g. artwork images), among others. 
The semantic lifting of such a variety of resources can be a serious bottleneck for the project activities. 
Several languages have been developed to either engineer content transformation (e.g. RML) or extending the SPARQL query language to access non-RDF resources (e.g. SPARQL Generate).
However, existing solutions require Semantic Web practitioners to learn a mapping language, or even combine multiple languages, for example requiring to use XPath for XML transformations.
In addition, these require Semantic Web practitioners to know the details of the original format (e.g. XML) as well as the target domain ontology.

In this paper, we don't propose a new language.
Instead, we aim at reducing the effort of Semantic Web practitioners in dealing with heterogeneous data sources by providing a generic, domain-independent meta-model 
as a \textit{facade} to wrap the original resource and to make it query-able \textit{as-if} it was RDF.
Specifically, we contribute a meta-model and associated algorithm for accessing non-RDF resources as RDF: \textit{Facade-X}.
Our approach can be implemented by overriding the SERVICE operator and does not require to extend the SPARQL syntax.
We compare our approach with the state of art methods RML and SPARQL Generate, and show how our solution has lower learning demands and cognitive complexity, and it is cheaper to implement and maintain,
while having comparable extensibility and efficiency (in our naive implementation).

In the next section we analyse the key requirements, building also on the work of~\cite{lefranccois2017sparql}.
In Section~\ref{sec:approach} we describe our approach for adopting \textit{facades} for re-engineering resources into RDF and give a formal definition of \textit{Facade-X}.
Section~\ref{sec:implementation} is dedicated to the prototype implementation of the approach in a software named SPARQL Anything.
Related work is discussed in Section~\ref{sec:relatedwork}.
We compare our approach with state of art methods (RML and SPARQL Generate) in Section~\ref{sec:evaluation}, before concluding our paper in Section~\ref{sec:conclusions}.
%

%% file: sections/requirements.tex
The motivation for researching  novel ways to transform non-RDF resources into RDF comes from the scenarios under development in the EU H2020 project SPICE: Social Cohesion, Participation, and Inclusion for Cultural Engagement. 
In this project, a consortium of eleven partners collaborate in developing novel ways for engaging with cultural heritage, relying on a \textit{linked data} network of resources from museums, social media, and businesses active in the cultural industry.
However, the majority of resources involved are not exposed as Linked Data but are released, for example, as CSV, XML, JSON files, or combinations of these formats.
In addition, the research activity aims at the design of task-oriented ontologies, producing multiple semantic \textit{viewpoints} on the resources and their metadata.
It is clear how the effort required for transforming resources could constitute a significant cost to the project.
In the absence of a strategy to cope with this diversity, content transformation may result in duplication of effort and become a serious bottleneck.
Table~\ref{tab:requirements} provides a summary of the requirements. 
\begin{table}[h]
    \caption{Requirements}
    \label{tab:requirements}
    \centering
    \begin{tabular}{|@{~}l|@{~}p{8cm}|}\hline
\textbf{Requirement} & \textbf{Description}\\\hline\hline
\tt{Transform} & Transform several sources having heterogeneous formats\\\hline
\tt{Query} & Query resources having heterogeneous formats\\\hline
\tt{Binary} & Support the transformation of binary formats\\\hline
\tt{Embed} & Support the embedding of content in RDF\\\hline
\tt{Metadata} & Support the extraction of metadata embedded in files\\\hline
%
\tt{Low learning demands} & Minimise the tools and languages that need to be learned\\\hline
\tt{Low complexity} & Minimise complexity of the queries\\\hline
\tt{Meaningful abstraction} & Enable focus on data structures rather than implementation details\\\hline
\tt{Explorability} & Enable data exploration without premature commitment to a mapping, in the absence of a domain ontology.\\\hline
\tt{Workflow} & Integrate with a typical Semantic Web engineering workflow\\\hline
\tt{Adaptable} & Be generic but flexible and adaptable\\\hline\hline
\tt{Sustainable} & Inform into a software that is easy to implement, maintain, and does not have evident efficiency drawbacks\\\hline
\tt{Extendable} & Support the addition of an open set of formats\\\hline
    \end{tabular}
\vspace{-0.8cm}
\end{table}
The main requirement is the ability to support users in transforming existing non-RDF resources having heterogeneous formats (\tt{Transform}).
In addition, the solution should be able to support cases in which practitioners only need to interrogate the content (\tt{Query}).
A valid approach should be able to cope with binary resources as well as textual formats (\tt{Binary}).
In the cultural heritage domain, metadata files are typically associated to repositories of binary content such as images in various formats. 
Applications may need to transfer data and metadata in a single operation, embedding the binary content in a data value (\tt{Embed}) and extracting metadata (\tt{Metadata}) from the file (from EXIF annotations).
%
\begin{sloppypar}We consider requirements related to usability and adoption. 
The approach should ideally limit the number of new languages and tools that need to be learned in order to transform and use non-RDF resources (\tt{Low learning demands}). 
This can be expected to both encourage adoption and reduce the learning curve for new users. 
The code that the user is required to develop in order to access the resources should be as simple as possible (\tt{Low complexity}). 
The approach should provide the user with a meaningful level of abstraction, enabling them to focus on the the structure of the data (e.g. data rows and hierarchies) rather than the details of how the structure has been implemented (\tt{Meaningful abstraction}). 
The approach should support an exploratory way of working in which the user does not have to prematurely commit to a domain ontology before they come to understand the data representation that they require (\tt{Explorability}). 
The resulting technology should be easily combined with typical Semantic Web engineering workflow (\tt{Workflow}). 
This requirement, already mentioned in~\cite{lefranccois2017sparql}, is interpreted considering that the solution should rely as much as possible on already existing technologies typically used by our domain users.
The approach should allow for a technical solution that is generic but easily \tt{Adaptable} to user tasks, for example, supporting symbol manipulation, variable assignments, and data type manipulation.\end{sloppypar} 

Finally, we look into requirements of software engineering.
The approach should be \tt{Sustainable} and inform a software that is easy to implement on top of existing Semantic Web technologies, easy to maintain, and does not have efficiency drawbacks compared to alternative state of the art solutions.
Ultimately, the system should be easy to extend (\tt{Extendable}) to support an open ended set of formats. 
%
%
%
%

%% file: sections/approach.tex
We introduce a novel approach to interrogate non-RDF resources with SPARQL.
Our opinion is that the task of transforming resources into RDF should be decoupled in two very different operations: (a) re-engineering, and (b) remodelling.
We define \textit{re-engineering} as the task of transforming resources minimising domain considerations, focusing on the meta-model. 
Instead, remodelling is the transformation of \textit{domain knowledge},  
where the original domain model is re-framed into a new one, whose main objective is to add semantics.
From this perspective, we propose to solve the re-engineering problem automatically and delegating the remodelling to the RDF-aware user.
\textit{How to use RDF to access heterogeneous source formats?}
We rely on the notion of \textit{facade} as \textit{"an object that serves as a front-facing interface masking more complex underlying or structural code"}\footnote{The Facade Design Pattern: \url{https://en.wikipedia.org/wiki/Facade\_pattern} (accessed 15/12/2020).}.
Applied to our problem, a facade acts as a generic meta-model allowing (a) to inform the development of transformers from an open ended set of formats, and (b) to generate RDF content in a consistent and predictable way.
In what follows, we describe a generic approach that can be used to develop facade-based connectors to heterogeneous file formats.
After that, we introduce Facade-X, which is the first of these interfaces, and describe how our facade maps to RDF. 
Finally, we design a method to inject facades into SPARQL engines. 
To support the reader, we introduce a guide scenario reusing the data of the Tate Gallery collection, published on GitHub\footnote{Tate Gallery collection metadata: \url{https://github.com/tategallery/collection}.}.
The repository contains CSV tables with metadata of artworks and artists and a set of JSON files with details about each catalogue record, for example, with the hierarchy of archive subjects. 
The file \texttt{artwork\_data.csv} includes metadata of the artworks in the collection such as \tt{id}, \tt{artist}, \tt{artistId}, \tt{title}, \tt{year}, \tt{medium}, and references two external resources: a JSON file with the artwork \tt{subjects} headings and a link to a JPG \tt{thumbnail} image.
Our objective is to serve this content to the Semantic Web practitioners for exploration and reuse.
\subsection{Resources, data sources, and facades}
In this section we give a formal definition of the three components of our approach: resources, data sources, and facades. In addition, we describe a generic algorithm for applying these concepts for re-engineering resources in RDF.

We consider a \textit{resource} anything accessible from a URL and distinguish it from its content, that we name \textit{data source}. The file \texttt{artwork\_data.csv}\footnote{Available at \url{https://raw.githubusercontent.com/tategallery/collection/master/artwork\_data.csv}} and the image $N04858\_8.jpg$ are resources and the CSV and JPG content are data sources.
We assume that a resource contains at least one data source\footnote{In principle, a resource may include multiple data sources, for example, an Excel spreadsheet may include several sheets.}. 
A data source can be named with the URL or have a different name\footnote{Although resources and data sources can be named with the same string (URL), we consider them different entities in our model.}.
We introduce the following predicates and associated axioms, in predicate logic:
\begin{lstlisting}[style=plogic]
$Resource(r)$ $DataSource(ds)$ $Name(n)$ $includes(r,ds)$ $hasName(ds,n)$
$\forall~r.Resource(r)\to\exists~ds.includes(r,ds)\land~DataSource(ds)$
$\forall~ds.DataSource(ds)\to\exists~n.hasName(ds,n)\land~Name(n)$
$\forall~ds.DataSource(ds)\to\exists~r.Resource(r)\land~includes(r,ds)$
$\forall~ds_1~\forall~ds_2~\forall~n~\forall~r.$
    $includes(r,ds_1)\land~includes(r,ds_2)\land~hasName(ds_1,n)\land~hasName(ds_2,n)$
        $\to~ds_1=ds_2$
$\forall~ds~\forall~n_1~\forall~n_2.hasName(ds,n_1)\land~hasName(ds,n_2)\to~n_1=n_2$
\end{lstlisting}
In addition, we refer to two additional concepts: RDF Graph and RDF Dataset,  as specified by RDF 1.1~\cite{rdf11}.
We now describe a generic algorithm for applying facades to resources and obtain RDF datasets capable of answering a certain query.
Let $Q$ be the set of all possible queries, $G$ the set of all possible graphs, $N$ the set of all possible graph names, $R$  the set of all possible resources and $DS$ the set of all data sources. 
We define: 
\begin{enumerate*}[label=\textit{(\roman*)}]

\item $D$ as a collection of named graphs (i.e. $D\subseteq N \times G$);

\item $A$ (i.e. the algorithm) as a function that given a resource ($r \in R$), a facade ($f \in F$), and a query ($q \in Q$), returns a collection of named graphs (i.e. one of the possible subsets of $N \times G$);

\item $F$ is a set of functions where each $f \in F$ associates a data source ($ds \in DS$) and a query ($q \in Q$) with a graph $g \in G$.
\end{enumerate*}
$A$ and $F$ can be formally defined as follows:
\begin{align*}
A: R \times F \times Q \to 2^{N \times G} && F=\{f|f: DS \times Q \to G\}
\end{align*}
Additionally,  given a query $q \in Q$, a resource $r\in R$ and its data sources $ds \in DS$, we define:
\begin{enumerate*}[label=\textit{(\roman*)}]

    \item $g_{ds,q}^* \in G$ as the graph which contains the minimal (optimal) set of triples required to answer $q$ on  $ds$;
    
    
    
    \item $D_{r,q}^* =\{ (n, g_{ds,q}^*) | includes(r,ds) \allowbreak  \text{ and }\allowbreak n \in N \text{  and } g_{ds,q}^* \in G $\} as the collection of minimal set of triples required to answer $q$ on $r$.
\end{enumerate*}
It is worth noticing that given a query and a resource neither $A$ nor any $f \in F$ has to return an optimal response (i.e. $D_{r,q}^*$ and $g_{ds,q}^* $), but they can return any super set of the optimum (i.e. any $g \in G \text{ such that } g_{ds,q}^* \subseteq g$).
We don't make any commitment on the underlying implementation of the facade with respect to the resource/data sources, apart from assuming that the resulting dataset will be sufficient, but not necessarily optimal, for answering the query.
\subsection{Facade-X}
We base the design of Facade-X on the distinction between containers and values.
Specifically, we define a \textit{container} as a set of uniquely identifiable \textit{slots}, each one of them including either another container or a data \textit{value}. 
Slot identifiers (\textit{keys}) can be either XSD strings ($StringKey$) or XSD positive integers ($NumberKey$). 
The predicate $Key$ is a reification of either an integer or a string, while the predicate $Value$ reifies a string only.
Containers can optionally be qualified by a \textit{type}. 
In Facade-X, data sources are referred to as \textit{root} containers. 
We specify our facade in predicate logic as follows:
\begin{lstlisting}[style=plogic]
$Root(c0)$ $Container(c1)$ $Slot(s1)$ $Key(n)$ 
$StringKey(n)$ $NumberKey(n)$ $Value(v1)$ $Type(t)$
$\forall k. StringKey(k) \rightarrow Key(k)$
$\forall k. NumberKey(k) \rightarrow Key(k)$
$\neg\exists k. NumberKey(k) \land StringKey(k)$
$\forall~c.Root(c) \rightarrow Container(c)$
\end{lstlisting}
In addition, we define relations between the model components, including definitions of domain and range:
\begin{lstlisting}[style=plogic]
$hasSlot(c,s)$ $hasType(c, t)$ $hasKey(s,k)$
$hasContainer(s,c)$ $hasValue(s,v)$
$\forall(x,y). hasSlot(x,y) \rightarrow Container(x) \land Slot(y)$
$\forall(x,y). hasType(x,y) \rightarrow Container(x) \land Type(y)$
$\forall(x,y). hasKey(x,y) \rightarrow Slot(x) \land Key(y)$
$\forall(x,y). hasContainer(x,y) \rightarrow Slot(x) \land Container(y)$
$\forall(x,y). hasValue(x,y) \rightarrow Slot(x) \land Value(y)$
\end{lstlisting}
We define a set of axioms describing additional properties of the meta-model. 
Only containers can have a type (but they don't have to), and there can only be one root container.
A slot can have either one container or one value and cannot have both.
A slot can be member of one container only and slots of a container are uniquely identified by their key:
\begin{lstlisting}[style=plogic]
$\forall(x,y). Root(x) \land Root(y) \rightarrow x = y$
$\neg\exists(x,y,z). hasContainer(x,y) \land hasValue(x,z)$ 
$\forall(x,y,z). hasContainer(x,y) \land hasContainer(x,z) \rightarrow y = z$
$\forall(x,y,z). hasValue(x,y) \land hasValue(x,z) \rightarrow y = z$
$\forall(x,y,z). hasSlot(x,y) \land hasSlot(z,y) \rightarrow x = z$
$\forall(c,s1,s2p,n). hasSlot(c,s1) \land hasSlot(c,s2) \land hasKey(s1,n) \land hasKey(s2,n)\rightarrow s1 = s2$
\end{lstlisting}
The data from our guide scenario can be represented as follows:
\begin{lstlisting}[style=plogic]
$Root(ds)$ 
$StringKey(id)$ $StringKey(artist)$ $StringKey(artistId)$ $StringKey(title)$
$hasSlot(ds,s1)$ $hasKey(s1,IntegerKey(1))$ $hasContainer(s1,r1)$ 
    $hasSlot(r1,r1s1)$ $hasKey(r1s1,id)$ $hasValue(r1s1,"1035")$
    $hasSlot(r1,r1s2)$ $hasKey(r1s2,artist)$ $hasValue(r1s2,"Blake~Robert")$
    $hasSlot(r1,r1s3)$ $hasKey(r1s3,artistId)$ $hasValue(r1s3,"38")$
    $hasSlot(r1,r1s4)$ $hasKey(r1s4,title)$ $hasValue(r1s4,"A~Figure~Bowing~...")$  [...]
\end{lstlisting}
Finally, we define mapping rules to RDF, where properties are built using string keys and resources can be either blank nodes or named IRIs\footnote{The system may allow users to define their own namespace, or reuse the name of the $ds$, and leave to the underlying machinery to mint IRIs.}:
\begin{lstlisting}[style=plogic]
$Root(ds) \stackrel{f}{\to} Triple(Resource(ds), rdf:type, fx:Root)$
$hasSlot(c,s) \land hasKey(s,k) \land StringKey(k) \land hasContainer(s,c1)$
    $\overset{f}{\to} Triple(Resource(c), Property(k), Resource(c1))$
$hasSlot(c,s) \land hasKey(s,k) \land IntegerKey(k) \land hasContainer(s,c1)$
    $\overset{f}{\to} Triple(Resource(c), ContainerMembershipProperty(k), Resource(c1))$
$hasSlot(c,s) \land hasKey(s,k) \land StringKey(k) \land hasValue(s,v)$
    $\overset{f}{\to} Triple(Resource(c), Property(k), Literal(v))$
$hasSlot(c,s) \land hasKey(s,k) \land IntegerKey(k) \land hasValue(s,v)$
    $\overset{f}{\to} Triple(Resource(c), ContainerMembershipProperty(k), Literal(v))$
$hasType(c,t) \overset{f}{\to} Triple(Resource(c), rdf:type, Resource(t))$
\end{lstlisting}
\todo{Clarify namespaces used, difference between vocabulary and SERVICE IRI handler.}
\todo{Those triples will be mapped to a named ds in the dataset.}
Our model maps into an RDF that mixes \textit{lists}, \textit{type} statements, and \textit{key-value} pairs.
Recent work suggests good practices for developing lists in RDF that are efficient to query~\cite{data2019modelling,dagasequential}, favouring container membership properties over nested structures to represent lists.
We define two namespaces, one for the primitive entity \tt{Root} and another for minting properties from keys\footnote{Not all strings are valid IRI local names. Implementations will need to apply heuristics to cope with corner cases in CSV or JSON keys.}. The above mappings produce the following \textit{Facade-X RDF}, from our example scenario:
\begin{lstlisting}[style=sparql]
@prefix fx: <http://sparql.xyz/facade-x/ns/>.
@prefix rdf: <http://www.w3.org/1999/02/22-rdf-syntax-ns#>.
@base <http://sparql.xyz/facade-x/data/>.
[] a fx:Root ;
   rdf:_1 [:id "1034"; :artist "Blake Robert"; :artistId "38"; ... 
   rdf:_2 [:id "16216"; :artist "Williams Terrick" :artistId "2149"; ... 
   rdf:_3 [:id "12086"; :artist "Pissarro Lucien" :artistId "1777"; ... 
   ...
\end{lstlisting}
\subsection{Using facades in SPARQL} 
%
The algorithm in Section 3.1 requires as input a URL and returns an RDF dataset as output.
We propose to \textit{overload} the SPARQL SERVICE operator by defining a custom URI-schema, based on the protocol \tt{x-sparql-anything:}, which is intended to behave as a \textit{virtual} remote endpoint. 
The related URI-schema supports an open-ended set of parameters specified by the facade implementations available. 
Options are embedded as key-value pairs, separated by comma.
Implementations are expected to either guess the source type from the resource locator or to obtain an indication of the type from the URI schema, for example, with an option "mime-type":
\begin{lstlisting}[language=sparql,basicstyle=\scriptsize,mathescape=true]
x-sparql-anything:mime-type=application/json; charset=UTF-8,location=$\dots$
\end{lstlisting}
Following our example scenario, users can write a query and select metadata from the CSV file, as well as embed the content of remote JPG thumbnails in the RDF. Multiple SERVICE clauses may integrate data from more files, for example, the JSON with details about artwork subjects. 
We leave the content of the CONSTRUCT section to be filled by the ontology engineer:
\begin{lstlisting}[style=sparql]
PREFIX fx: <http://sparql.xyz/facade-x/ns/>
PREFIX xyz: <http://sparql.xyz/facade-x/data/>
PREFIX rdf: <http://www.w3.org/1999/02/22-rdf-syntax-ns#>
CONSTRUCT {
    [...] # Amazing ontology here
} WHERE {
    BIND (IRI(CONCAT(STR(tate:), "artwork-", ?id )) AS ?artwork) .
    BIND (IRI(CONCAT(STR(tate:), "artist-", ?artistId )) AS ?artist) .
    SERVICE <x-sparql-anything:csv.headers=true,location=file:./artwork_data.csv> {
        []  xyz:id ?id ;              xyz:artist ?artistLabel ;
            xyz:accessionId ?accId ;  xyz:artistId ?artistId ;
            xyz:title ?title;         xyz:medium ?medium ;
            xyz:year ?year ;          xyz:thumbnailUrl ?thumbnail .
    }
    # JPEG Thumbnail from the Web
    BIND (IRI(CONCAT("x-sparql-anything:location=", ?thumbnail )) AS ?embedJPG ).
    SERVICE ?embedJPG { [] rdf:_1 ?imageInBase64 }.
    # JSON File with subjects
    BIND (IRI(CONCAT("x-sparql-anything:file:./artworks/", ?accId )) AS ?subJSON ).
    SERVICE ?subJSON { [ xyz:id ?subjectId ; xyz:name ?subjectName ] }. 
}
\end{lstlisting}\vspace{-0.5cm}

%% file: sections/implementation.tex
In this section we describe SPARQL Anything which is meant to provide a proof-of-concept of our approach. 
SPARQL Anything implements a stack of transformers mapped to media types and file extensions. 
The framework allows the addition of an open-ended set of transformers as Java classes.
During execution, a query manager intercepts usage of the SERVICE operator and in case the endpoint URI has the \tt{x-sparql-anything} protocol, it parses the URI extracting the resource locator and parameters.
Default parameters are: \texttt{mime-type}, \texttt{locator}, \texttt{namespace} (to be used when defining RDF resources), and \texttt{root} (to use as the IRI of the root RDF resource, instead of a blank node), and \texttt{metadata}. SPARQL Anything will project an RDF dataset during query execution including the data content and optionally a graph named \tt{http://sparql.xyz/facade-x/data/metadata}, including file metadata extracted from image files (also in Facade-X). Specific formats may support specific parameters. For example, the Text triplifier supports a regular expression to be used by a tokenizer that splits the content in a list of strings (defaults to the space character).
Similarly, the CSV triplifier\todo{Should we change "triplifier" with "transformer"?} allows to specify whether to use the first row as headers or only use column indexes.
More information on the currently supported formats can be found in the project page\footnote{\url{http://github.org/sparql-anything/sparql-anything}.}.

We validated the generality of Facade-X as a meta-model with relation to the triplifiers currently implemented in SPARQL Anything. 
We already considered CSV in the guide example.
The following JSON example, also derived from the Tate Gallery open data, can be mapped to our model as in the associated listing.
\begin{lstlisting}[style=plogic]
{ "fc": "Kazimir Malevich", 
  "id": 1561, 
  "places": [
    { "name": "Ukrayina", "type": "nation" }, 
    { "name": "Moskva, Rossiya", "type": "inhabited_place" }
  ], 
  "url": "http://www.tate.org.uk/art/artists/kazimir-malevich-1561" }

$Root(malevic)$
$StringKey(fc) StringKey(id) StringKey(places) StringKey(name)$
$StringKey(type) StringKey(url)$
$NumberKey(1) NumberKey(2)$
$hasSlot(malevic, s_{fc}) \land hasKey(s_{fc},fc) \land hasValue(s_{fc}, "Kazimir Malevich")$
$hasSlot(malevic, s_{id}) \land hasKey(s_{id},id) \land hasValue(s_{id}, 1561)$
$hasSlot(malevic, s_{places}) \land hasKey(s_{places},places) \land hasContainer(s_{places}, c_{places})$
$hasSlot(c_{places}, s_{place/1}) \land hasKey(s_{place/1},1) \land hasContainer(s_{place/1}, ukraina)$
$hasSlot(ukraina, s_{ukr/name}) \land hasKey(s_{ukr/name},name) \land$     
    $hasValue(s_{ukr/name}, "Ukrayina")$
$hasSlot(ukraina, s_{ukr/type}) \land hasKey(s_{ukr/type},type) \land hasValue(s_{ukr/type}, "nation")$
$hasSlot(c_{places}, s_{place/2}) \land hasKey(s{place/2},2) \land hasContainer(s_{place/2}, moskva)$
$hasSlot(moskva, s_{mos/name}) \land hasKey(s_{mos/name},name) \land$ 
    $hasValue(s_{mos/name}, "Moskva, Rossiya")$
$hasSlot(moskva, s_{mos/type}) \land hasKey(s_{mos/type},type) \land$       
    $hasValue(s_{mos/type}, "inhabited\_place")$
\end{lstlisting}

%
Finally, we show how an excerpt from a catalogue record in XML, can be interpreted with our Facade-X (this also applies to HTML):
\begin{lstlisting}[style=plogic]
  <OGT hint="OGGETTO">
    <OGTD hint="Definizione">reperti antropologici ...</OGTD>
    <OGTT hint="Tipologia">reperto osteo-dentario</OGTT>
    ...

$Root(ogt) \land hasType(record, "OGT")$
$Container(ogtd) \land hasType(ogtd,"OGTD")$
$Container(ogtt) \land hasType(ogtd,"OGTT")$
$StringKey(hint)$
$NumberKey(1) NumberKey(2)$
$hasSlot(ogt, s_{1}) \land hasKey(s_{1},1) \land hasValue(s_{1}, ogtd)$
$hasSlot(ogt, s_{2}) \land hasKey(s_{2},2) \land hasValue(s_{2}, ogtt)$
$hasSlot(ogt, s_{hint}) \land hasKey(s_{hint},hint) \land hasValue(s_{hint}, "OGGETTO")$
...
\end{lstlisting}



%% file: sections/relatedwork.tex
Related work includes semantic web approaches to content re-engineering,
approaches to extending the functionalities of SPARQL, and
research on end-user development and human interaction with data. 

\begin{sloppypar}In ontology engineering, non-ontological resource re-engineering refers to the process of taking an existing resource and transforming it into an ontology~\cite{villazon2010pattern}. 
These family of approaches integrate resource transformation within the methodology, where domain knowledge plays a central role. 
Triplify~\cite{auer2009triplify} is one of the first tools aiming at converting sources into RDF in a domain independent way. The approach is based on mapping HTTP URIs to ad-hoc database queries, and rewriting the output of the SQL query into RDF. 
Other tools are based on the W3C Direct Mapping recommendation~\cite{w3c:directmapping} for relational databases.
Systems are available for automatically transforming data sources of several formats into RDF (Any23\footnote{\url{http://any23.apache.org/}}, JSON2RDF\footnote{\url{https://github.com/AtomGraph/JSON2RDF}}, CSV2RDF\footnote{\url{http://clarkparsia.github.io/csv2rdf/}} to name a few). A recent survey lists systems to lift tabular data~\cite{fiorelli2021lifting}. While these  tools have a similar goal (i.e. enabling the user to access the content of a data source as if it was in RDF), the (meta)model used for generating the RDF data  highly depends on the input format. 
All these approaches are not interested in the requirement of providing a common useful abstraction to heterogeneous formats.
A long history of mapping languages for transforming heterogeneous files into RDF can be considered superseded by RML~\cite{dimou2014rml}, including a number of approaches for ETL-based transformations~\cite{arenas2021knowledge}.
We consider RML as representative of general data integration approaches such as OBDA~\cite{xiao2018ontology}.
This family of solutions are based on a set of declarative mappings. 
The mapping languages incorporate format-specific query languages (e.g. SQL or XPath) and require the practitioner to have deep knowledge not only of the input data model but also of standard methods used for its processing.
Recent work acknowledges how these languages are built with machine-processability in mind~\cite{heyvaert2018declarative} and how defining or even understanding the rules is not trivial to users.\end{sloppypar}

We survey approaches to extend SPARQL. A standard method for extending SPARQL is by providing custom functions\footnote{ARQ provides a library of custom functions for supporting aggregates such as computing a standard deviation of a collection of values. ARQ functions: \url{https://jena.apache.org/documentation/query/extension.html} (accessed 15/12/2020).}, or by using so-called \textit{magic} properties. 
This approach defines custom predicates to be used for instructing specific behaviour at query execution. 
SPARQL-Generate~\cite{lefranccois2017sparql} introduces a novel approach for performing data transformation from heterogeneous sources into RDF by extending the SPARQL syntax with new operators~\cite{lefranccois2017sparql}:
GENERATE, SOURCE, and ITERATOR.
Custom functions perform ad-hoc operations on the supported formats, for example, relying on XPath or JSONPath. 
Other approaches extend SPARQL \textit{without} changes to the standard syntax.
BASIL~\cite{daga2015basilar} allows to define parametric queries by enforcing a convention in SPARQL variable names. As a result, SPARQL query templates can be processed with standard query parsers.
SPARQL Micro-service~\cite{michel2019enabling} provides a framework that, on the basis of API mapping specification, wraps web APIs in SPARQL endpoints and uses JSON-LD profile to translate the JSON responses of the API into RDF. In this paper, we follow a similar, minimalist approach and extend SPARQL by \textit{overriding} the behaviour of the SERVICE operator. 
We compare our proposal with SPARQL Generate and RML in detail in the evaluation section.

%
%
%

Motivation\todo[author=Paul]{If short of space maybe this could be removed. I could just reference a couple of key papers related to abstraction and explorability in the requirements section} for our work resides in research on end-user development and human interaction with data.
End-user development is defined by~\cite{lieberman2006end} as \textit{"methods, techniques, and tools that allow users of software systems, who are acting as non-professional software developers, at some point to create, modify or extend a software artefact"}.  
Many SPARQL users fall into the category of end-user developer. 
In a survey of SPARQL users,~\cite{warren2018using} found that although 58\% came from the computer science and IT domain, other SPARQL users came from non-IT areas, including social sciences and the humanities. 
Findings in this area~\cite{panko2010revising} suggest that the data with which users work is more often primarily list-based and/or hierarchical rather than tabular. 
For example,
~\cite{hall2019lish} proposes an alternative formulation to spreadsheets in which data is represented as \textit{list-of-lists}, rather than tables. 
Our proposal goes in this direction and accounts for recent findings in end-user development research.

%% file: sections/evaluation.tex
We conduct a comparative evaluation of SPARQL Anything with respect to the state of art methods RML and SPARQL Generate.
First, we analyse in a quantitative way the cognitive complexity of the frameworks.
Second, we conduct a performance analysis of the reference implementations.
Finally, we discuss the approaches in relation to the requirements elicited in Section~\ref{sec:requirements}. 
Competency questions, queries, experimental data, and code used for the experiment are available on the GitHub repository of the SPARQL Anything project\footnote{\url{https://github.com/spice-h2020/SPARQL Anything/tree/main/experiment}}.
\vspace{-0.4cm}
\subsubsection{Cognitive Complexity Comparison.}\label{sec:evaluation:complexity}
We present a quantitative analysis on the cognitive complexity of SPARQL Anything, SPARQL Generate and RML frameworks. 
One effective measure of complexity is the number of distinct items or variables that need to be combined within a query or expression \cite{halford2004development}. Such a measure of complexity has previously been used to explain difficulties in the comprehensibility of Description Logic statements \cite{warren2015making}. 
Specifically, we counted the number of tokens needed for expressing a set of competency questions.
We selected four JSON files from the case studies of the SPICE project where each file contains the metadata of artworks of a collection.
Each file is organised as a JSON array containing a list of JSON objects (one for each artwork).
This  simple data structure avoids favouring one approach over the others.
Then, an analysis of the schema of the selected resources allowed us to define a set of 12 competency questions (CQs) that were then specified as SPARQL queries or mapping rules according to the language of each framework, in particular:
\begin{enumerate*} [label=(\roman*)]
    
    \item 8 CQs (named q1-q8), aimed at retrieving data from the sources, were specified as SELECT queries (according to SPARQL Anything and SPARQL Generate);
    
    \item  4  CQs (named q9-q11), meant for transforming the source data to RDF, were expressed as CONSTRUCT queries (according to SPARQL Anything and SPARQL Generate) or as mapping rules complying with RML. 
    These queries/rules intend to generate a blank node for each artwork  and to attach the artwork's metadata as dataproperties of the node.
    
\end{enumerate*}
Finally, we tokenized the queries (by using "()\{\},;{}\textbackslash{}n\textbackslash{}t\textbackslash{}r{\ttfamily\char32} as token delimiters) and we computed the total number of tokens  and the number of distinct tokens needed for each queries.
By observing the average number of tokens per query we can conclude that RML is very verbose  (109.75 tokens) with respect to SPARQL Anything (26.25 tokens) and SPARQL Generate (30.75 tokens) whose verbosity is similar (they differ of the $\sim$6.5\%).
However, the average number of \textit{distinct} tokens per query shows that SPARQL Anything requires less cognitive load than other frameworks.
In fact, while SPARQL Anything required 18.25 distinct tokens, SPARQL Generate needed 25.5 distinct tokens ($\sim$39.72\% more) and RML 45.25 distinct tokens ($\sim$150\% more).

\vspace{-0.4cm}
\subsubsection{Performance Comparison.}
\begin{figure}[t]
     \centering
     \begin{subfigure}[b]{0.45\textwidth}
         \centering
         \includegraphics[width=\textwidth]{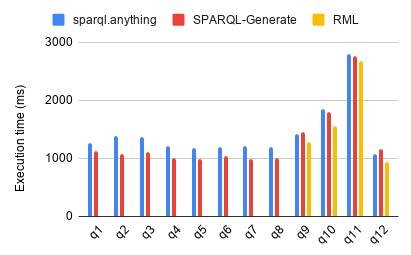}
         \caption{Execution time per  query.}
         \label{fig:performance_comparison_increasing_query}
     \end{subfigure}
     \hfill
     \begin{subfigure}[b]{0.54\textwidth}
         \centering
         \includegraphics[width=\textwidth]{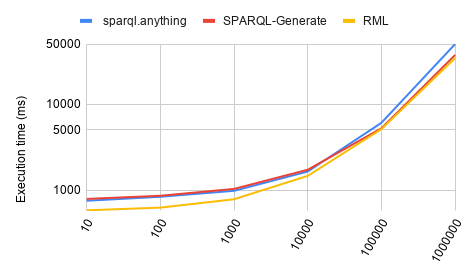}
         \caption{Execution time with increasing input size.}
         \label{fig:performance_comparison_increasing}
     \end{subfigure}
    \caption{Analysis of the the execution time.}
    \label{fig:performance_comparison}\vspace{-0.5cm}
\end{figure}

We assessed the performance of three frameworks in generating RDF data.
All of the tests described below were run three times and the average time among the three executions is reported.
The tests were executed on a MacBook Pro 2020 (CPU: i7 2.3 GHz, RAM: 32GB).
Figure~\ref{fig:performance_comparison_increasing_query} shows the time needed for evaluating the SELECT queries q1-q8 and for generating the RDF triples according to the CONSTRUCT queries/mapping rules q9-q12.
The three frameworks have comparable performance.
We also measured the performance in transforming input of increasing size. 
To do so, we  repeatedly concatenated the data sources in order to obtain a JSON array containing 1M JSON objects and we cut this array at length 10, 100, 1K, 10K and 100K.
We ran the query/mapping q12 on these files and we measured the execution time shown in Figure~\ref{fig:performance_comparison_increasing}.
We observe that for inputs with size smaller than 100K the three frameworks have equivalent performance. 
With larger inputs, SPARQL Anything is slightly slower than the others. 
The reason is that, in our naive implementation, the data source is completely transformed and loaded into a RDF dataset in-memory, before the query is evaluated. 
However, implementations could \textit{stream} the triples during query execution, or transform the optimal triple set for the query solution, thus achieving better performance on large input. However, we leave this optimisations to future work. 

\vspace{-0.1cm}
\begin{table}[t]
    \centering
    \caption{Formats supported by RML, SPARQL Generate, and SPARQL Anything.}
    \label{tab:formats}
    \begin{tabular}{c|cccccccccc}\hline
             & JSON & CSV & HTML & Bin. & XML & RDB & Text & Embed & Meta. & Spread.  \\\hline\hline
             
        \textbf{RML}  & 
        \checkmark & 
        \checkmark & 
        \checkmark & 
          & 
        \checkmark  & 
        \checkmark & 
        \checkmark & 
          & 
          & 
        \\
        
         \textbf{SPARQL-Generate}  & 
         \checkmark & 
         \checkmark & 
         \checkmark & 
         \checkmark  &
         \checkmark & 
         \checkmark & 
         \checkmark  & 
          & 
         & 
        \\
        
         \textbf{SPARQL Anything}  & 
        \checkmark & 
         \checkmark & 
         \checkmark & 
         \checkmark &
         \checkmark & 
          & 
         \checkmark   & 
         \checkmark & 
         \checkmark 
         & 
         \checkmark 
        \\
    \end{tabular}
\vspace{-0.5cm}
\end{table}
\subsubsection{Requirements satisfaction and discussion}
We discuss the requirements introduced in Section~\ref{sec:requirements}.

\tt{Transform}, \tt{Binary}, \tt{Embed}, and \tt{Metadata}. All the frameworks support users in  transforming  heterogeneous formats with few differences (a comparison is provided in Table~\ref{tab:formats}).
Currently, SPARQL Anything and SPARQL-Generate cover the largest  set of input formats.
SPARQL-Generate however does not support embedding content (\tt{Embed}) and extracting metadata from files (\tt{Metadata}). 
Both features are not supported by RML, which doesn't support plain text as well. 
SPARQL Anything allows users to query spreadsheets, but it is not  able to handle relational databases yet\footnote{However, relational tables can be mapped using an approach similar to CSV and spreadsheets tables. A dedicated component is currently being developed.}.
SPARQL Anything is the only tool supporting the extraction of metadata and the embedding of binary content.

\tt{Query}. In terms of query support, while RML requires data to be transformed first and then uploaded to a SPARQL triple store, SPARQL Anything and SPARQL-Generate enable users to query resources directly. 

%
%
\tt{Low learning demands}. SPARQL Generate uses an extension to SPARQL 1.1 to transform source formats into RDF. RML provides an extension to the R2RML vocabulary in order to map source formats into RDF. Therefore either a SPARQL extension or a new mapping language has to be learned to perform the translation. In the case of Facade-X, no new language has to be learned as data can be queried using existing SPARQL 1.1 constructs.

\tt{Low complexity}. Complexity can be measured as the number of distinct items or variables that need to be combined with the query. In experiments, Facade-X is found to perform favourably in comparison to SPARQL Generate and RML.

\tt{Meaningful abstraction}.
Differently from RML and SPARQL-Generate, which require users to be knowledgeable of the source formats and their query languages (e.g. XPath, JSONPath etc.), Facade-X users can access a resource \textit{as if} it was an RDF dataset, hence the complexity of the non-RDF languages is completely hidden to them. The cost for this solution is limited to the users which are required to \textit{explore} the facade that is generated and tweak the configuration via the Facade-X IRI schema.


\tt{Explorability}. With SPARQL Generate and RML, the user needs to commit to a particular mapping or transformation of the source data into RDF. However, the data representation required to carry out a knowledge intensive task often emerges from working with data and cannot be wholly specified in advance (this is a crucial requirement of our project SPICE). 
By distinguishing the processes of re-engineering and re-modelling, Facade-X enables the user to avoid prematurely committing to a mapping and rather focus on querying the data within SPARQL, in a domain-independent way. 

\tt{Workflow}. All the technologies considered can in principle be integrated with a typical Semantic Web engineering workflow. However, while we cannot assume that Semantic Web experts have knowledge of RML, XPath, and SPARQL Generate, we can definitely expect knowledge of SPARQL.

\tt{Adaptable}. All technologies provide a flexible set of methods for data manipulation, sparql.aything relying on plain SPARQL. We make the assumption that SPARQL itself is enough for manipulating variables, content types, and RDF structures. It is an interesting, open research question to investigate content manipulation patterns in the various languages and compare their ability to meet user requirements. 
%

\tt{Extendable} and \tt{Sustainable}.
Our approach can be implemented within existing SPARQL query processors with minimal development effort. Extending SPARQL Anything requires to write a component that exposes a data source format as Facade-X. Facade-X does not need to be encoded in the software but serves as a reference for mapping an open ended set of formats. 
In contrast, extending SPARQL Generate and RML requires extending the user toolkit to handle the specificity of the formats, exposing to users new functions for querying, filtering, traversing, and so on.
In addition, our approach leads to a more sustainable codebase. To give evidence of this statement, we use the tool cloc\footnote{cloc: \url{https://github.com/AlDanial/cloc} (accessed 15/12/2020).} to count the lines of Java code required to implement the core module of SPARQL Generate in Apache Jena (without considering format-specific extensions\footnote{For SPARQL Generate, we only considered the code included in the submodule sparql-generate-jena.}) and the RML implementation in Java\footnote{RMLMapper: \url{https://github.com/RMLio/rmlmapper-java}.}.
SPARQL Generate and RML require developing and maintaining 12280 and 7951 lines of Java code, respectively.
We developed the prototype implementation of SPARQL Anything with 3842 lines of Java code, including all the currently supported transformers.

%% file: sections/conclusions.tex
In this paper, we presented an opinionated approach for making non-RDF resources query-able with SPARQL.
We contributed a general approach to apply \textit{facades} to content re-engineering and a specific instance of this approach, Facade-X, which defines a general meta-model akin to a \textit{list-of-lists}.
We compared our approach with the state of art methods RML and SPARQL Generate and demonstrated how our solution has lower learning demands and cognitive complexity, and it is cheaper to implement and maintain, while having comparable extensibility.
Next, we will extend the range of supported formats of SPARQL Anything, including relational databases, Microsoft Office files, and binary content other then images, and develop new strategies for performance optimisation.
Moreover, we will perform a user study for investigating the cognitive implications of using Facade-X as a meta-model with respect to arbitrary RDF, and compare the tools in terms of expressivity and ability to meet user requirements.
Finally, other \textit{facades} can be designed as well.
It is an interesting research question to investigate content manipulation patterns in alternative facades and evaluate their benefit for content exploration and transformation.